\documentclass{article}

\usepackage{arxiv}

\usepackage[utf8]{inputenc} 
\usepackage[T1]{fontenc}    
\usepackage{hyperref}       
\usepackage{url}            
\usepackage{booktabs}       
\usepackage{amsfonts}       
\usepackage{nicefrac}       
\usepackage{microtype}      
\usepackage{lipsum}		

\hypersetup{
	colorlinks=true, 
	citecolor=blue} 

\title{Modeling failures times with dependent renewal type models via exchangeability}
\date{May 12, 2019\\
to appear in \textit{Statistics: A Journal of Theoretical and Applied Statistics}}	

\author{
  Arrigo Coen\thanks{CORRESPONDING AUTHOR: Arrigo Coen, Email: coen@ciencias.unam.mx} \\
  Departamento de Matem\'aticas, Facultad de Ciencias \\
  Universidad Nacional Aut\'onoma de M\'exico\\
  M\'exico, CDMX, Apartado Postal 20-726, 01000, M\'exico\\
  \texttt{coen@ciencias.unam.mx} 
   \And
 Luis Guti\'errez \\
  Departamento de Estad\'istica, Pontificia Universidad Cat\'olica de Chile\\
  Millennium Nucleus Center for the Discovery of Structures in Complex Data\\
   Santiago, C\'odigo Postal 7820436, Chile\\
  \texttt{llgutier@mat.uc.cl} \\
   \AND
   Rams\'es H. Mena \\
   Departamento de Probabilidad y Estad\'istica, Instituto de Investigaciones en Matem\'aticas Aplicadas y en Sistemas\\
   Universidad Nacional Aut\'onoma de M\'exico, M\'exico\\
   CDMX, Apartado Postal 20-726, 01000, M\'exico\\
   \texttt{ramses@sigma.iimas.unam.mx} \\
}

\usepackage{amssymb,amsmath,amsthm,amsfonts,epsfig}

\usepackage{varioref}
\usepackage{hyperref}
\usepackage{cleveref}

\usepackage{subfig}
\usepackage{dsfont} 
\usepackage{color,soul} 

\newcommand{\iid}{\stackrel{\rm iid}{\sim}}

\newtheorem{thm}{Theorem}
\newtheorem{prop}{Proposition}
\newtheorem{defi}{Definition}

\newcommand{\mb}{\mathbb}

\newcommand{\prob}[1]{\mathbb{P}\left[#1\right]}
\newcommand{\esp}[1]{\mathbb{E}\left[#1\right]}

\newcommand{\clg}{\mathcal} 
\def\simiid{\stackrel{\mbox{\scriptsize{iid}}}{\sim}}

\newcommand{\Fvar}{F} 
\newcommand{\Qrand}{\tilde{F}}
\newcommand{\espmu}[1]{\mathbb{E}_\mu\left[#1\right]}
\newcommand{\dist}[1]{\mathrm{#1}} 

\begin{document}
\maketitle

\begin{abstract}
	Failure times of a machinery cannot always be assumed independent and identically distributed, e.g.  if after reparations the machinery is not restored to a same-as-new condition. Framed within the renewal processes approach, a generalization that considers exchangeable inter-arrival times is presented. The resulting model provides a more realistic approach to capture the dependence among events occurring at random times, while retaining much of the tractability of the classical renewal process. Extensions of some classical results and special cases of renewal functions are analyzed, in particular the one corresponding to an exchangeable sequence driven by a Dirichlet process. The proposal is tested through an estimation procedure using simulated data sets and with an application to the reliability of hydraulic subsystems in load-haul-dump machines.
\end{abstract}


\section{Introduction}\label{intro}

In reliability engineering, the time to failure of a component, as well as, the expected number of failures in a time horizon is crucial in the planning of the production in manufacturing lines. Depending on the complexity of the machinery, the times between failures or component replacements cannot always be assumed independent and identically distributed (i.i.d.)
(cf. \cite{Reliability,Rausand2004}). Hence, the classical renewal processes theory
(e.g., \cite{Parzen_Stochastic_B99}) is not always adequate. In particular, the independence assumption is violated if, upon reparation, the machinery is not restored to a same-as-new condition. Similarly, the equality in distribution assumption is rarely satisfied, though relaxing it typically requires multiple realizations of the failure process.

Indeed, dependent renewal-type models are increasingly demanded due to their versatility and wide applicability. Applications include: occurrence of rare events \cite{Segerdahl_a70}, streams of customers  \cite{Kuczura_73}, lifetimes modeling \cite{Mercer_Wear_Dependent_Renewal_a61, Shanthikumar_a87}, disease activity \cite{Cook_Two-state_mixed_renewal} and web applications \cite{Liu06}, among others. Theoretical contributions in some specific dependent renewal-type models, including Markov renewal processes, can be found in \cite{Shanthikumar_General_shock_models_a83,Zamba11,Adekpedjou_a13,Markov_renewal_Asmussen_16}. For a classical account of the theory of Markov renewal processes, we refer the reader to \cite{Markov_Renewal_Cinlar_a75}. 

While some of these models provide with excellent extensions, most apply only for specific failure distributions. Clearly, allowing for certain dependence, while retaining  flexibility in the choice of failure distribution, imposes a serious mathematical-applicability tradeoff. A good mathematical compromise and natural step to relax the i.i.d. assumption, keeping the failure distribution flexibility, is to make use of well known distributional symmetries for the joint distribution of the failure times. Among these, the most tractable is exchangeability \cite{Finetti1931, deFinetti1937}, which in this context is equivalent to say that all failures are dependent in a similar magnitude but conditional independent given the overall uncertainty of failures has been resolved or quantified. {Examples of these types of renewal epochs, with a common factor that generates the dependence, could be seen in} \cite{william2002}.

Formally, a sequence of $\mathbb{R}_+$-valued random variables,  $T=\{T_i:i\in\mb{N}\}$,  is said to be exchangeable if for any $ n\in\mb{ N} $ and every permutation $ \pi $ of $ \{1,\ldots,n\} $, the vector $ \left(T_{\pi(1)}, \ldots, T_{\pi(n)}\right) $ has the same distribution as $ \left(T_1,\ldots,T_n\right)$.  It is easy to see that ${\mathsf{Corr}}(T_i,T_j)\geq 0$ and it is the same for all $i\neq j$ with $i,j\in\mb{N}$. {This is in accordance with reliability applications, 
	as positive dependence is commonly observed among renewal times} \cite{Rausand2004}{.} The beauty of  exchangeability  is evident from {the de Finetti representation theorem} (see, e.g., \cite{Hewitt_Symmetric_measures_A55}), which states that   $T$  is exchangeable, if and only if, for any $n\geq 0$, and any Borel sets $\{A_i: i=1,\dots,n\}$, 
there exists a probability measure $\mu$ on the set of probability measures on $\mathbb{R}_+$, $\clg{F}$, such that
\begin{eqnarray*}
	\mathbb P(T_1\in A_1,\ldots, T_n\in A_n)=\int_\clg{F} F(A_1)\cdots F(A_n)\mu(d F).
\end{eqnarray*}

The measure $\mu$  characterizes the exchangeable sequence $\{T_i: i\in\mb{ N}\} $ and it is known as the de Finetti's measure driving $T$. A commonly used representation of de Finetti's theorem is through the hierarchical form 
\begin{eqnarray*} 
	T_{i}\mid F &\iid & F, \qquad  {i\in\mb{ N}}, \\
	F &\sim & \mu, \nonumber
\end{eqnarray*}which disentangles the conditional i.i.d. property.

Comprehensibly, exchangeable failure times would not cover all possible dependence scenarios, however we claim that it poses an excellent alternative to the i.i.d. case without compromising its tractability, thanks to the aforementioned conditional independence property. Notice that the i.i.d. case is recovered when de Finetti's measure degenerates in a particular failure distribution, i.e. when $\mu=\delta_F$, with $\delta_a$ denoting a unit point mass at $a$. The concept of exchangeability to generalize renewal-type processes was previously touched by \cite{Huang_Point_Processes_with_the_Exchangeable_a90}, where  exchangeable renewals are used to characterize the class of {mixed renewal processes} with the Markov property. Here, we further elaborate on such a class of \textbf{$\mu$-mixed renewal processes}, in particular we exploit the uniqueness of the de Finetti's measure for infinite exchangeable sequences as well as some of the more recent Bayesian ideas to construct them. This substantially adds to \cite{Huang_Point_Processes_with_the_Exchangeable_a90} proposal, in particular it allows us to study some concrete examples.

The remainder of the manuscript is organized as follows: In Section \ref{sect:properties}, we give some definitions and properties of the mixed renewal function together with its connections to the classical i.i.d. case. This section also includes a discussion of the induced correlation structure and confronts it with that of a non-homogeneous Poisson process. Section \ref{sect:modeling} examines various important cases corresponding to specific choices of exchangeable sequences. In particular, we discuss a model where the dependence is driven by setting de Finetti's measure to be the Dirichlet Process \cite{Ferguson73}. An inferential strategy for mixed renewal processes is presented  and illustrated in Section \ref{sect:inference}. The illustrations include an application in the context of reliability. We conclude the manuscript in Section \ref{sect:discussion}, with a discussion and future directions.  The proofs are deferred to the Appendix.

\section{Exchangeable mixed renewal processes}\label{sect:properties}

In the classical i.i.d. renewals case, the renewal function characterizes completely the distributional properties of the renewal process. Such characterization does not  follow for the exchangeable case. Indeed, having dependence among failure times precludes from a characterization via marginal properties, and thus a generalized version of such renewal function is required. That said, 
given the conditional i.i.d. property of exchangeable sequences, the resulting renewal function nicely connects with classical renewal theory.  In what follows we formalize the class of exchangeable mixed renewal processes and study some of its properties.

\begin{defi}\label{def:ERP}
	Let $ T=\{T_i:i\in\mb{N}\}$ be a sequence of nonnegative exchangeable random variables with de Finetti's measure $\mu$, and marginal distribution $ F$, such that $F(0)<1$. We say $ S_{\mu}=\{S_n:n\in\mb{N}\} $ is a \textbf{$\mu$-mixed adding process} if 
	\[ S_0=0\qquad S_n=T_1+T_2+\cdots+T_n\quad n\ge1.\]
	Accordingly, we define the \textbf{$\mu$-mixed renewal process} $ N_{\mu}=\{N(t):t\ge0\} $  by \[ N(t) =\sup\{n:S_n\le t \}, \quad t\ge0.\]
\end{defi}

\begin{defi}\label{def:renewal_fun}
	The \textbf{$\mu$-mixed renewal function} is given by
	\begin{equation}\label{eq_Ut_como_Esp_de_N}
	U(t)=\esp{N(t)}, \quad  t\ge0.
	\end{equation}
\end{defi}

For example, if the exchangeable sequence, $T$, is defined by assuming $T_i\mid \theta\sim\mathsf{Exp}(\theta)$ and $\theta\sim\mathsf{U}(0,2\lambda)$, one can easily see that $U(t)=\lambda t$ with the marginal interarrival density given by $f_{T_i}(t)=t^{-2}\left[1-e^{-2\lambda t}(1+2\lambda t)\right]$. Thus, while in the i.i.d. having $U(t)=\lambda t $ reduces to $\mathsf{Exp}(\lambda)$ interarrivals times, in the exchangeable case such uniqueness is clearly not satisfied. Furthermore, if instead we have $T_i\mid \delta\sim\mathsf{Ga}(2, {{\delta} })$ and $f(\delta) = \alpha k^\alpha \delta^{-1-\alpha} \mathds{1}_{ \delta\ge k}$, i.e. a Pareto distribution with scale parameter $k>0$ and shape parameter $\alpha>0$, then one has that $U(t)=\infty$ when $\alpha\leq 1$ and $U(t)<\infty$, when $\alpha> 1$. Here, we will only consider those cases satisfying $U(t)<\infty$ for all $t\geq 0$, which are the cases of interest in most practical applications.  This, clearly imposes some conditions on the exchangeable sequence $T$.

With these observations and definitions at hand, we can now study the connections between  $\mu$-mixed renewal processes and the classical i.i.d. renewal case. 
\begin{defi}\label{def:conditionals_renewal_functions}
	Let $\mathcal{F}$ as before and $\mathfrak{F}$ the corresponding Borel $\sigma$-algebra induced by the topology of weak convergence. For all $ t\ge0 $ and $ \mathcal{G}\in \mathfrak{F}$, we define the \textbf{conditional renewal set function} as
	\begin{equation*}
	U\left(t\mid \tilde{F} \in\mathcal{G}\right) = \int_{\mathcal{G}}  \esp{N(t) \mid \tilde{F}=F}\mu(d F) 
	\end{equation*}
	and the  \textbf{conditional renewal function} as
	\begin{equation*}
	U(t\mid F)=\esp{N(t)\mid \tilde{F}=F},\qquad  F\in \mathcal{F}.
	\end{equation*}
\end{defi}

Notice that  if the set $\mathcal{G}$ has only one random element, e.g. $\mathcal{G}=\left\{F\right\}$ then $U(t\mid \tilde{F} \in\mathcal{G})$ and $U(t\mid F)$ coincide. When the measure $\mu$ is degenerated to a single element, $U(t\mid F)$ is equal to the classical renewal function with renewals distributed as $F$. In what follows, when the de Finetti's measure degenerates in a parametric family of distributions, $\mathsf{F}_\Theta:=\{F_\theta: \theta\in \Theta\}$, i.e. when $\mu\left(\mathsf{F}_\Theta\right)=1$, we use the notation $U(t\mid \theta)$ $ :=U(t|F_\theta) $ to refer to the conditional renewal function. The following results formalize the connection between the conditional renewal function and the $\mu$-mixed renewal function.  The proofs are deferred to the Appendix.

\begin{thm}\label{theo:relationU(t)andU(t|F)}
	The relation between the conditional renewal function and the $\mu$-mixed renewal function is given by
	\[  U(t)=\int_\mathcal{F} U(t\mid F)\mu(d F), \qquad t\ge0.\]
\end{thm}

Theorem \ref{theo:relationU(t)andU(t|F)} shows an  obvious, yet very useful, connection between the independent and exchangeable cases. The renewal function, in the exchangeable case, is a weighted average of independent renewal functions, where the weight is given through the de Finetti's measure. As a consequence of this last relation, one could extend various classical results. In particular, we have the following relevant observations.
\begin{prop}\label{prop_U_sum_conv_F}
	If $U(t)$ is a $\mu$-mixed renewal function, then
	\[ 	 U(t)=\sum_{n=1}^{\infty}\int_\mathcal{F} \Fvar^{*n}(t) \mu(d\Fvar),\qquad t\ge0, \]
	where $ \Fvar^{*n} $ denotes the $n$-fold convolution of $ F $.
\end{prop}

As we will see, in the concrete cases studied in Section \ref{sect:modeling}, Proposition \ref{prop_U_sum_conv_F} gives a useful representation of the mixed renewal function. Other results are also at hand, e.g. if $U(t)$ is a $\mu$-mixed renewal function, then one can deduce that 
\begin{equation}\label{eq_cota_inf_Ut}
U(t)\ge \int_\clg{D}\dfrac{t}{\esp{T_1|F}}\mu(dF)-1,\qquad t\ge0, 
\end{equation}
where  $\clg{D}=\left\{Q\in\clg{F}:\esp{T_1|Q}<\infty \right\}$.

\begin{prop}\label{prop_Laplace_ex_renewal}
	The Laplace transform of the $\mu$-mixed renewal function is given by
	\begin{equation*} 
	L_U(s)=\int_\mathcal{F} \dfrac{L_{\Fvar}(s)}{1-L_{\Fvar}(s)}\mu(d\Fvar),\qquad s\ge0,
	\end{equation*}
	where $ L_{F} $ is the Laplace transform of the distribution $F$.
\end{prop}

\begin{thm}\label{thm_first_eq_of_U}
	The $\mu$-mixed renewal function and the conditional renewal function have the relation 
	\begin{equation*}
	U(t) =\espmu{\Qrand(t)}+\espmu{\Qrand*U(t|\Qrand)}.
	\end{equation*}
\end{thm} This latter result helps to generalize the concept of proper renewal equations to the exchangeable case. Let us remember that in the classical, non-delayed renewal process,    $A(t)=\esp{f(N(t))}$ satisfies the linear integral equation $A(t)=a(t)+(F*A)(t)$.  

\begin{thm}\label{thm_key_ex_rev_theorem}
	Let $ a(t) $ be a bounded positive function over bounded intervals. Then there is one and only one solution to the equation $ A(t, \clg{G}):\mb{R}^+\times  \mathfrak{F} \to\mb{R}^+ $, where
	\begin{equation}\label{eq_renewal} 
	A(t,\clg{G})=a(t)+ \espmu{\Qrand*A(t,\{\Qrand\})\mathds{1} _{\Qrand\in\clg{G}}}, \qquad t\ge 0, \clg{G}\in \mathfrak{F},
	\end{equation}
	with 
	\[ \sup_{0\le s\le t, \clg{G}\in \mathfrak{F}}|A(s,\clg{G})|<\infty, \]
	and the solution is given by
	\begin{equation*}
	A(t,\clg{G})=a(t)+ \espmu{a*U(t|\Qrand)  \mathds{1} _{\Qrand\in\clg{G}}}, \qquad t\ge0,\clg{G}\in \mathfrak{F}. 
	\end{equation*}
	We refer to the solution of {\eqref{eq_renewal}} as a \textbf{general mixed renewal process}. 
\end{thm}

Theorem \ref{thm_key_ex_rev_theorem} provides a new characterization of a general $\mu$-mixed renewal process, and as a byproduct establishes the dependence  structure that the process will follow. For example, when $\mathcal{G}=\mathcal{F}$, and setting $A(t):=A(t,\mathcal{F})$, \eqref{eq_renewal} simplifies to
\begin{equation}\label{eq:solution_2}
A(t) =a(t)+\espmu{\Fvar*A(t,\{\Qrand\})},\qquad t\ge0.
\end{equation}

The time evolution of $A$ is driven by two terms: $a(t)$, which  modulates the general tendency of the process, and the second term, which relates the general mean of the process to the local mean behavior, when conditioning to the value of the random measure $\tilde{F}$. Adjusting these two components might result  appealing when aiming at different interactions among them.

To illustrate this result assume that the exchangeable sequence $T$ is defined via  $T_i\mid \theta\sim\mathsf{Exp}(\theta)$ and $\theta\sim\eta$, and consider the following two scenarios: (i) $\eta=\sum_{i=1}^m p_i \delta_{\alpha_i}$ and (ii) $\eta=\mathsf{Ga}(\alpha,\lambda)$, with fixed $ \lambda,\alpha,\alpha_i>0 $ and $ p_i\in(0,1)$ such that $ \sum p_i=1 $. In other words, discrete (i) and continuous (ii) mixtures of the exponential distribution with conditional mean $1/\theta$. We further assume the drift function is given by $a(t)=1-e^{-\beta t}$, for $t\geq 0$. Thus applying  Theorem~\ref{thm_key_ex_rev_theorem}, one has that for (i)
\begin{eqnarray*}
	A(t)&=&a(t)+\mathbb{E}_\eta[a(t)*U(t\mid \theta)]\nonumber\\
	&=&1- e^{-\beta t}+\sum_{i=1}^mp_i\int_0^t \left(1-e^{-\beta (t-x)}\right)
	\alpha_idx\nonumber\\
	&=&1- e^{-\beta t}+\left(t+\dfrac{e^{-\beta t}-1}{\beta}\right)\sum_{i=1}^m p_i \alpha_i.
\end{eqnarray*}
For (ii), a similar application of Theorem~\ref{thm_key_ex_rev_theorem} leads to 
\begin{eqnarray*}
	A(t)=-\frac{\alpha }{\beta  \lambda }+e^{-t\beta} \left(\frac{\alpha }{\beta  \lambda }-1\right)+t+1.
\end{eqnarray*}This latter case corresponds to a  Pareto marginal  
renewal distribution given by
\begin{equation*} 
f_T(t) = \dfrac{\alpha \lambda^\alpha }{(t+\lambda)^{\alpha +1}} \mathds{1}_{[0,\infty)}(t), 
\end{equation*}with mean $\lambda/(\alpha-1)$, for $\alpha>1$. Figure \ref{solreneq} below, displays some scenarios where the difference between the i.i.d. (classical)  and the $\mu$-mixed renewal solutions exhibit a significant difference. For  both cases the marginals, $T_i$s, are the same, i.e. discrete mixture of the exponential distribution or Pareto. In other words, departing from an exchangeable assumption in the renewal sequence, does not compromise the i.i.d./classical tractability of the solutions, while it might exhibit big differences. 

\begin{figure}[!ht]
	\begin{center}
		\begin{tabular}{cc}
			{\centering \includegraphics[scale=0.72]{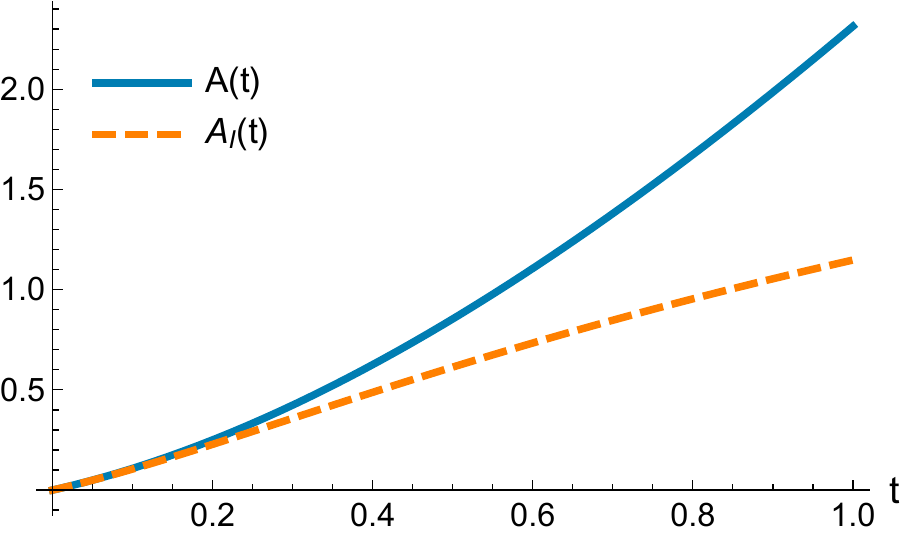}} &
			{\centering \includegraphics[scale=0.72]{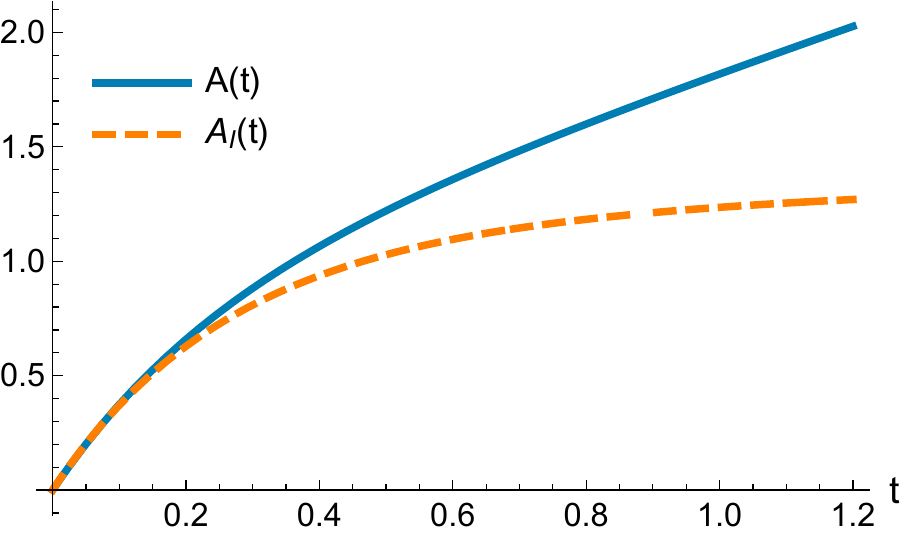}} \\
			{\centering (i)} &  {\centering (ii)}
		\end{tabular}
	\end{center}
	\caption{Solutions for the renewal equations for the discrete (i) and continuous (ii) mixtures of exponentials. Both cases, the i.i.d., $A_I$, and the exchangeable, $A$,  are displayed.  For  (i), $m=2$,  
		$(\alpha_1,\alpha_2)=(0.1, 10)$ and $(p_1,p_2)=(0.5,0.5)$. For (ii) $(\alpha,\beta,\lambda)= (2, 4, 3)$. In both cases we are assuming $\beta=0.9$.
		\label{solreneq}}
\end{figure} 

\subsection{Induced correlation structure}
The exchangeability  inherent to the $\mu$-mixed renewal process induces an appealing dependence structure. In general, the  covariance function can be computed using   
\begin{align}
\mathsf{Cov}\left(N(t),N(t+s)\right) =& \espmu{\mathsf{Cov}\left(N(t),N(t+s)|\Qrand \right)} \nonumber \\
&+  \mathsf{Cov}_{\mu}\left(\mathbb{E}{\left[N(t)|\Qrand\right]},\mathbb{E}{\left[N(t+s)|\Qrand\right]}\right). \label{cova}
\end{align}

This covariance could be compared with that corresponding to other type of counting processes. A popular choice, used  in the modeling of repairable systems (e.g. \cite{Lindqvist_2006}), is the non-homogeneous Poisson process, $\{N^{*}(t), t\geq 0\}$,  with intensity function $\lambda(t)$.  In such case,  one has $\mathsf{Cov}(N^*(t), N^*(t+s))=\Lambda(t)$, where $\Lambda(t)=\int_{0}^{t}\lambda(x)dx$. In this case, the correlation  simplifies as 
\begin{equation}\label{eq:corr1}
\mathsf{Corr}(N^*(t), N^*(t+s))=\left(\frac{\Lambda(t)}{\Lambda(t+s)}\right)^{1/2}.
\end{equation}

It is important to note that, being an independent increments process, the covariance itself characterizes the non-homogeneous Poisson process. In contrast,  one can find different  $\mu$-mixed renewal processes inducing a given 
covariance form. As observed from  Theorem \ref{thm_key_ex_rev_theorem}, a richer dependence structure is needed to characterize the $\mu$-mixed renewal process.  Indeed, a correlation function of the form  \eqref{eq:corr1}, can be obtained by a $\mu$-mixed renewal process.  Specifically, if $T_i\mid \theta\sim\mathsf{Exp}(\theta)$ and $ \theta\sim F $,  the $\mu$-mixed renewal processes has  covariance  given by 
\[{\mathsf{Cov}(N(t),N(t+s)) = t\mathbb{E}_{F}(\theta) + t(t+s)\mathsf{Var}_{F}(\theta)}.  \] 
Furthermore, if  $F$ is such that $\mathbb{E}(\theta)=\phi$ and $\mathsf{Var}(\theta)=\sigma^2$, the correlation function of the $\mu$-mixed renewal processes reduces to
\begin{equation}\label{eq:corr2}
\mathsf{Corr}(N(t), N(t+s))=\frac{t(\phi+(t+s)\sigma^2)}{\sqrt{t(t+s)}\sqrt{(\phi+t\sigma^2)(\phi+(t+s)\sigma^2)}}.
\end{equation} which has the same form as in \eqref{eq:corr1} with $\Lambda(t)=t/(\phi+t\sigma^2)$. In other words, the  above $\mu$-mixed renewal process recovers the same covariance structure as that of a non-homogeneous Poisson process with the integrable rate $\lambda(x)=\phi/(\sigma^2 x+\phi)^2$.

It is well known that some non-homogeneous Poisson processes, or even some Cox processes, coincide with some renewal processes (see, e.g. \cite{Yannaros1994}). However, these classes of counting processes do not fall under the category of $\mu$-mixed renewal process. Having said that, using a $\mu$-mixed renewal process, one can capture the same second order dependence properties of some Poisson random measures. 

\section{Modeling strategies with mixed renewal processes}\label{sect:modeling}
Here we explore various strategies making use of the results presented in  previous sections. In particular, we further elaborate on the  parametric and nonparametric constructions of the exchangeable renewals. 

\subsection{$\mu$ induced by a parametric construction}
In the previous sections, we have used a well-known Bayesian mechanism to construct exchangeable sequences, namely via a conditionally   i.i.d. sequence of random variables following a parametrized distribution, where the corresponding parameter(s) is then assigned another --prior-- distribution. More specifically, one could assume, for instance, that the exchangeable inter-arrivals sequence is conditionally given defined as   
\begin{eqnarray}\label{eq:parametricmodel1}
T_i\mid \lambda &\simiid & \mbox{Er}(m, \lambda),\quad i\in\mb{ N}\\ \nonumber
\lambda & \sim & \mbox{Ga}(\alpha, 1),
\end{eqnarray}

\noindent where $\mbox{Er}(m, \lambda)$ denotes an Erlang distribution with shape parameter $m\in \mb{N}$ and rate parameter $\lambda>0$. With this 
specification, the marginal distribution of $T_1$, thus of any $T_i$, is
\[f_{T_1}(t)=\frac{\Gamma(\alpha+m)\ t^{m-1}}{\Gamma(\alpha)\Gamma(m)(1+t)^{m+\alpha}},\] which is a particular case of a generalized beta distribution of the second kind  $\mbox{GB2}(T; a, b, p, q)$, where $a=b=1$, $p=m$ and $q=\alpha$ (see, \cite{generalization_beta_McDonald_95}). Also, for any $n\geq 1$, the joint density of $(T_{1},\ldots, T_{n})$ simplifies to
\begin{eqnarray*}
	f_{ T}(t_{1},\ldots,t_{n})=\frac{\Gamma(\alpha+nm) \left[ \prod_{i=1}^{n}t_{i} \right]^{m-1}}{\Gamma(\alpha)(m-1)!^{n}\left(1+\sum_{i=1}^{n}t_{i} \right)^{\alpha+nm}}.
\end{eqnarray*}It follows that, when $\alpha>2$, $ \esp{T_1}=m/(\alpha-1)$ and $\esp{T_1^{2}}=(m+1)\esp{T_1}/(\alpha-2)$ so 
\[
{\mathsf{Var}(T_1)}=\frac{m(\alpha+m-1)}{(\alpha-2)(\alpha-1)^{2}} \quad \mbox { and  } \quad {\mathsf{Corr}(T_{i}, T_{j})}=\frac{m}{\alpha+m-1}.
\]

Notice that for a fixed $m$, if $\alpha {\to} \infty$ then $ {\mathsf{Corr}(T_{i},T_{j}) \to} 0$. On the other hand, if $\alpha \downarrow 2$, then ${\mathsf{Corr}(T_{i},T_{j})\to} \frac{m}{m+1}$. Finally, for a fixed $\alpha$, if $m {\to} \infty$ then ${\mathsf{Corr}(T_{i},T_{j}) \to} 1$. That is, depending on the values of $m$ and $\alpha$, model \eqref{eq:parametricmodel1} is flexible  enough  to capture correlations between $0$ and $1$.

Given the known expression for the  conditional renewal function \eqref{eq:parametricmodel1} is 
\[U(t\mid \lambda) =  \frac{\lambda t}{m}+\frac{1}{m}\sum_{k=1}^{m-1}\frac{z^{k}}{1-z^{k}}(1-e^{-\lambda t (1-z^{k})}),\]
and applying Theorem \ref{theo:relationU(t)andU(t|F)}, it follows that 
the $\mu$-mixed renewal function is 
$$U(t)=\frac{\alpha t}{m}+\frac{1}{m}\sum_{k=1}^{m-1} \frac{z^{k}}{1-z^{k}}\left(1-\frac{1}{(1+t(1-z^{k}))^{\alpha}}\right),$$
where $z=e^{2\pi i/m}$. In particular, when $m=1$, i.e.  $T_i\mid \lambda \simiid \mbox{Exp}(\lambda)$, the marginal distribution of $T_1$ is given by 
\begin{eqnarray*}
	f_{T_1}(t)=\frac{\alpha}{(1+t)^{\alpha+1}},
\end{eqnarray*}  
which corresponds to a Pareto distribution with support in $[0,\infty)$. In this model,  $0< {\mathsf{Corr} (T_{i},T_{j})}<1/2$. Other quantities of interest in the mixed renewal process are $U(t)=\alpha t$, $ {\mathsf{Var}( N(t))}=\alpha t\left(1+ t \right)$ and $ {\mathsf{Cov}(N({t}),N({t+s}))}=\alpha t\left[ 1+(t+s)\right]$. Which confronted with 
$U(t\mid \lambda)=\lambda t$, ${\mathsf{Var}}(N({t})\mid \lambda)=\lambda t$ and ${\mathsf{Cov}}(N(t),N({t+s})\mid \lambda)=\lambda t$,  corresponding to a Poisson process, features some over-dispersion.

\subsection{A nonparametric Bayesian prior as choice for $\mu$}
Let us consider the exchangeable sequence for the inter-arrival times driven by  the  Dirichlet process \cite{Ferguson73}. Such process is crucial  in Bayesian nonparametric statistics (see, e.g. \cite{bayesian_nonparametrics_hjort2010}).  In other words, consider
\begin{eqnarray}\label{eq:nonparametric1}
T_i\mid F &\simiid & F,\\
F &\sim& \mbox{DP}(\alpha, H), \nonumber
\end{eqnarray}

\noindent where $\mbox{DP}(\alpha, H)$ denotes a Dirichlet process with precision parameter $\alpha>0$ and base distribution $H$.  The joint distribution of $(T_{1},\ldots,T_{n})$  can be factorized using the well-known P\'olya urn  predictive distribution, i.e. for any $n>1$ 
\begin{eqnarray*}
	T_n|T_{n-1},T_{n-2},\ldots,T_1 \sim \dfrac{\alpha}{\alpha+n-1}H + \dfrac{1}{\alpha+n-1}\sum_{i=1}^{n-1} \delta_{T_i},
\end{eqnarray*}
\noindent which weights between new and observed inter-arrival times. It easily follows that
\[ {\mathsf{Corr} (T_i,T_j)}= \frac{1}{\alpha+1},  \quad \mbox { for every }  i,j\in\mb{ N}. \] 

\begin{prop}\label{prop:nonparametric1}
	For the exchangeable renewals \eqref{eq:nonparametric1} we have
	\begin{enumerate}
		\item[(i)] For every fixed $ n\in\mb{ N} $ and $ H_j(t):=H(t/j)$,
		\[\prob{S_n\le t} = \dfrac{n!}{(\alpha)_n}\sum_v \left(\prod_{j=1}^n \dfrac{\alpha^{v_j}}{j^{v_j}v_j!}\right) \left( H_1^{*v_1}\ast H_2^{*v_1}\ast\cdots\ast H_n^{*v_n}\right) (t),\]
		the sum running over  vectors $ v=(v_1,\ldots,v_n) \in\mb{ N}$ satisfying $\sum_{j=1}^njv_j\!=\!n$.
		\item[(ii)] 
		$$ U(t) = \sum_{n=1}^\infty \dfrac{n!}{(\alpha)_n} \sum_v\left( \prod_{j=1}^n\dfrac{\alpha^{v_j}}{j^{v_j}v_j!}\right) \left( H_1^{*v_1}\ast H_2^{*v_1}\ast\cdots\ast H_n^{*v_n}\right) (t), $$
		where $(\alpha)_n=\alpha(\alpha+1)\cdots(\alpha+n-1)$.
	\end{enumerate}
\end{prop}

Model \eqref{eq:nonparametric1} together with the results of Proposition \ref{prop:nonparametric1} provide a general framework to define a nonparametric renewal mixed model. In fact, particular models can be defined selecting the base measure of the Dirichlet process. For instance, if $H(t):=1-e^{-\lambda t}$, i.e. an exponential distribution with mean $1/\lambda$, one verifies  that  $ H_{j}(t)= 1-e^{-\lambda t/j}$, leading  thus to
\[{H_{j}^{*v_{j}} }(t)=1-\sum_{u=0}^{v_{j}-1}\frac{1}{u!}e^{-\lambda t/j}(\lambda t/j)^{u},\]i.e. an Erlang distribution with parameters $\lambda/j$ and $v_{j}$. To obtain the distribution of $ S_n $ we  use the moment generation function of  $ {H_i^{*v_i} }$, given by
\begin{align*}
M_{S_n|V}(s|v) &= \prod_{i=1}^n\left(\dfrac{\lambda/i}{\lambda/i-s}\right)^{v_i}\\
&= \left(\prod_{m=1}^nm^{-v_m}\right) \left[\dfrac{1}{(1-t/\lambda)^{v_i}\times(1/2-t/\lambda)^{v_2}\times\cdots\times(1/n-t/\lambda)^{v_n}}\right]\\
&= \left(\prod_{m=1}^nm^{-v_m}\right) \sum_{i=1}^n\sum_{j=1}^{v_j} \dfrac{{ K_{i,j}^{(n)} } }{(1/i-t/\lambda)^j},
\end{align*}
where {$ K_{i,j}^{(n)} $} are the coefficients of the partial fractions. This allow us to obtain the conditional distribution 
\begin{equation*}
F_{S_n|V}(t|v)   = \left(\prod_{m=1}^nm^{-v_m}\right) \sum_{i=1}^n\sum_{j=1}^{v_j} {K_{i,j}^{(n)}} i^j\left[1-e^{-\lambda t/i}\sum_{u=0}^{j-1} \dfrac{(\lambda t/i)^u}{u!} \right],
\end{equation*}
which is a mixture of Erlang distributions. Adding over all possible $\sum_i^niv_i=n$, we obtain the distribution of $ S_n $ 
\[ F_{S_n}(t)   =\sum_v \left(\prod_{m=1}^nm^{-v_m}\right) \sum_{i=1}^n\sum_{j=1}^{v_j} {K_{i,j}^{(n)}} i^j\left[1-e^{-\lambda t/i}\sum_{u=0}^{j-1} \dfrac{(\lambda t/i)^u}{u!} \right].\]
Hence,  the $\mu$-mixed renewal function is given by 
\begin{equation*}
U(t) = \sum_{n=1}^\infty \sum_v\left(\prod_{m=1}^nm^{-v_m}\right) \sum_{i=1}^n\sum_{j=1}^{v_j} {K_{i,j}^{(n)}} i^j\left[1-e^{-\lambda t/i}\sum_{u=0}^{j-1} \dfrac{(\lambda t/i)^u}{u!} \right].
\end{equation*}
This result allow us to compute an approximation for $U(t)$. The advantage of using this equation lies in the fact that the partial fraction constants  {$ K_{i,j}^{(n)} $}  are easy to compute and that the convolutions $ F_{S_n}(t) $ converge quickly to zero for big values of $n$.

\section{Inference in mixed renewal processes}\label{sect:inference}
This section illustrates an inferential strategy for $\mu$-mixed renewal processes. In particular, for the applications we have in mind, we depart from the more general setting where we have various realizations of the arrival sequences, i.e.  within the framework of partial exchangeability. 
Hence, we consider $k$ sequences of $\mb{R}^+$-valued exchangeable random variables, with the $i$-th sequence denoted by  $T_{i1}, T_{i2},\ldots, T_{in_i}$.  We further assume that for $i=1,\ldots,k$, 
\begin{eqnarray*}
	T_{i1},\ldots, T_{in_i}\mid \lambda_i & \iid & \mbox{Er}(m, \lambda_i), \quad  \\ 
	\lambda_i & \iid & \mbox{Ga}(\alpha, 1).
\end{eqnarray*} Let $\boldsymbol t_i(n_i):=\{t_{i1},\ldots,t_{in_i} \}$, the joint density of the $k$ exchangeable sequences is 
\begin{eqnarray*}
	f_{\boldsymbol T}(\boldsymbol t_{1}(n_1),\ldots, \boldsymbol t_{k}(n_k))= \frac{\prod_{i=1}^k\Gamma(\alpha+n_i m)\left[\prod_{i=1}^k\prod_{j=1}^{n_i} t_{ij} \right]^{m-1}} { \Gamma(\alpha)^k \Gamma(m)^{\sum_{i=1}^k n_i}\prod_{i=1}^k \left[ 1+\sum_{j=1}^{n_i} t_{ij} \right]^{\alpha+n_i m} }.
\end{eqnarray*}

The maximum likelihood estimation for $\alpha$ and $m$ can be obtained via standard numerical optimization methods.

\subsection{Illustrations with simulated data}

\begin{figure}[!ht]
	\centering
	\subfloat[]{\includegraphics[scale=.45]{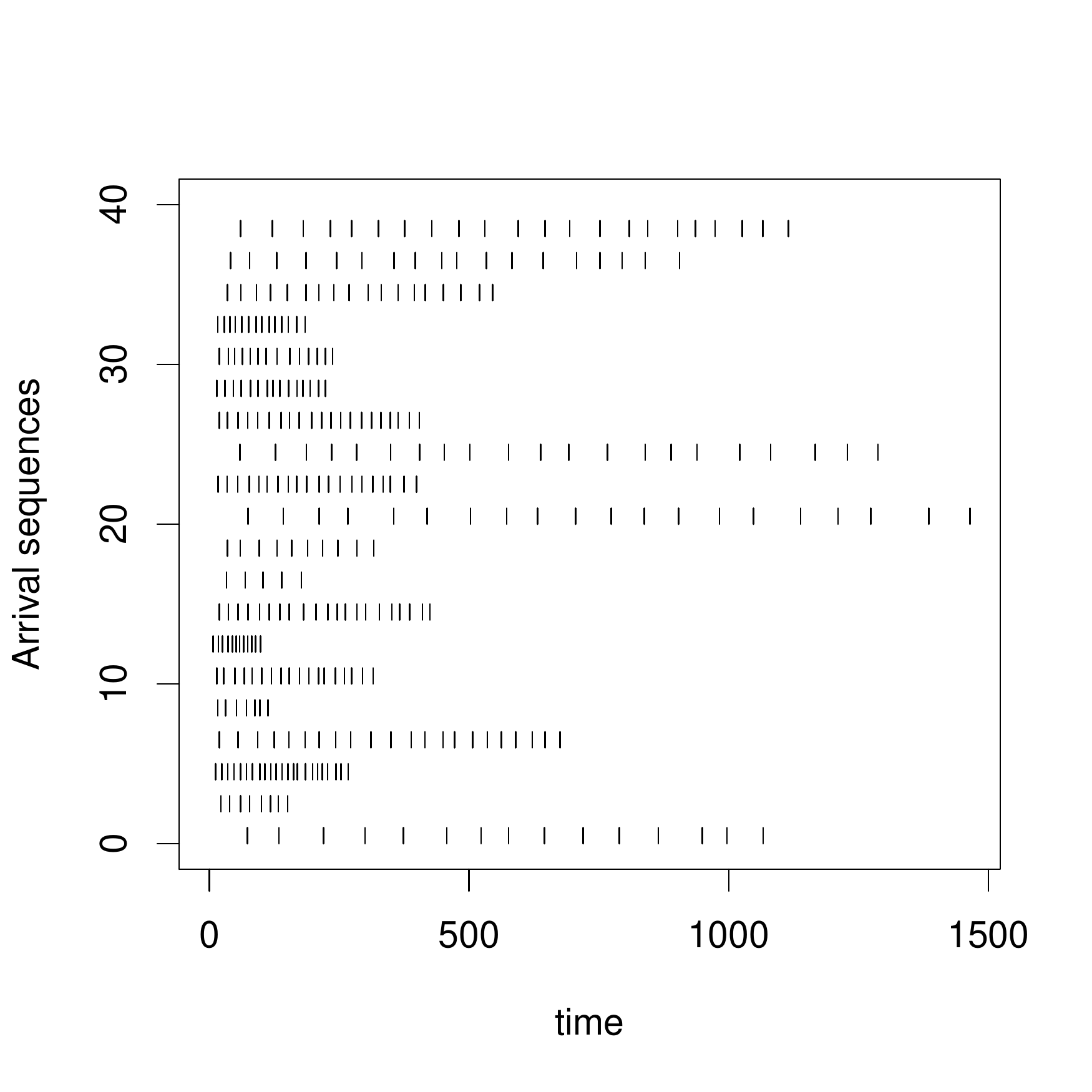}} 
	\subfloat[]{\includegraphics[scale=.45]{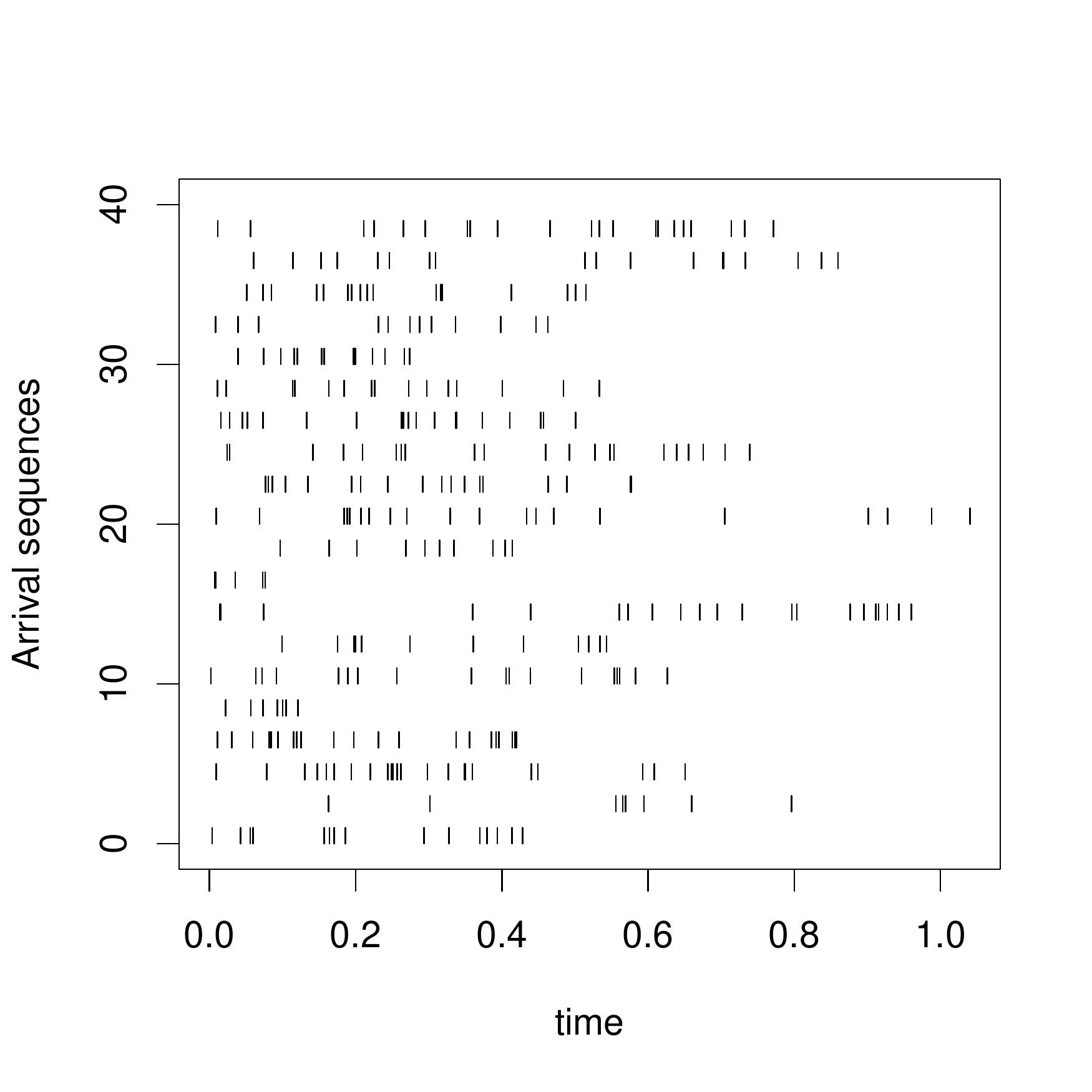}} \\
	\subfloat[]{\includegraphics[scale=.45]{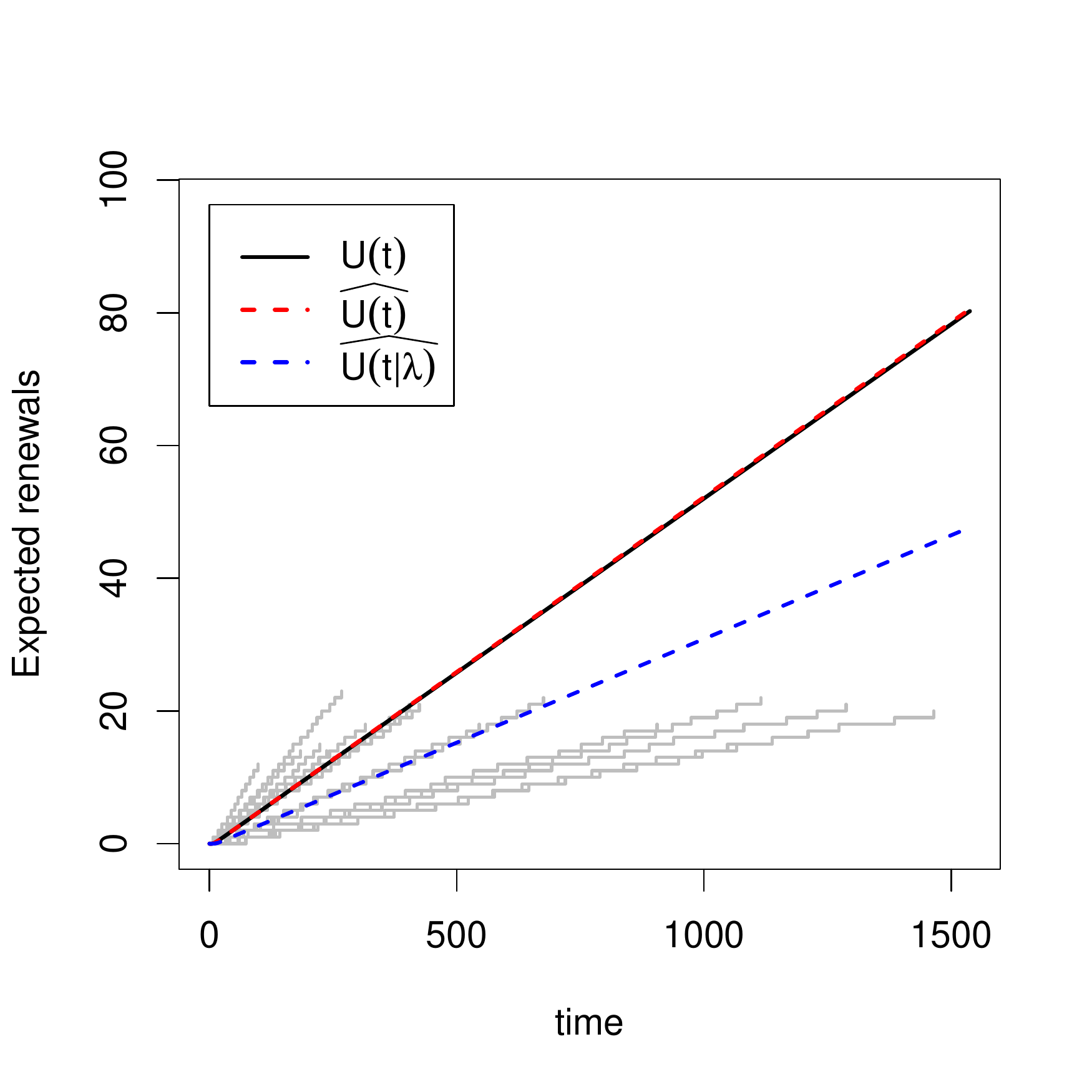}} 
	\subfloat[]{\includegraphics[scale=.45]{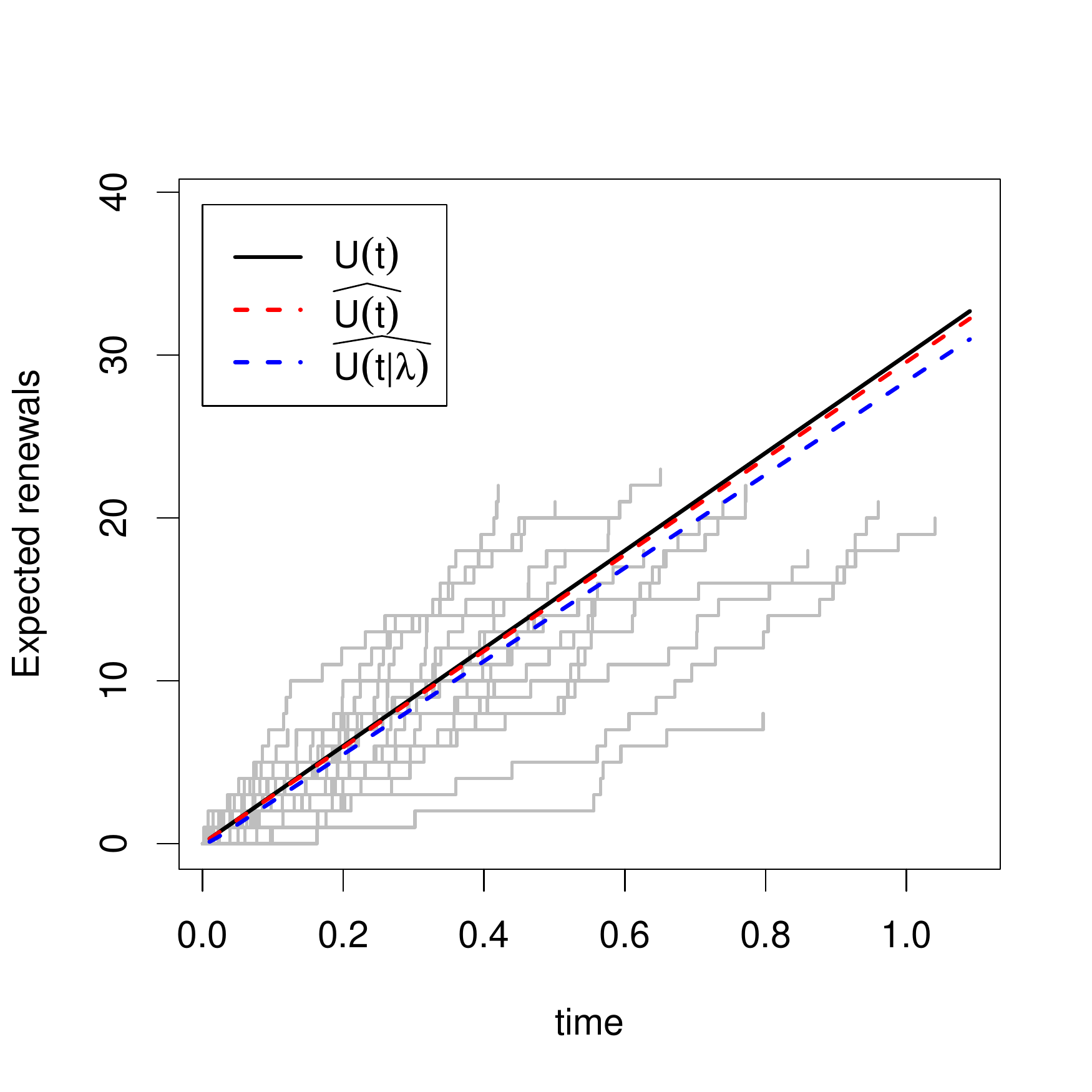}} 
	\caption{Arrival sequences of the data together with the true and estimated renewal functions. Left panels Example 1, right panels Example 2.}
	\label{fig:example12}
\end{figure}

We consider two examples, one with correlation close to one (Example 1) and the other one with correlation close to zero (Example 2). For each example, we simulated $k=20$ exchangeable sequences,  each one of dimension $n_i$. The values of $n_i$, $i=1,\ldots,20$ were fixed at: 
$15$, $8$, $23$, $22$, $7$, $18$, $12$, $21$, $5$, $10$, $20$, $20$, $21$, $21$, $15$, $14$, $14$, $18$, $18$ and $22$. Figure \ref{fig:example12} shows the simulated sequences {(subfigures (a) and (b))} and the true and estimated renewal functions {(subfigures (c) and (d)}). The true $\mu$-mixed renewal function denoted by $U(t)$ is represented by the black continuous line, the estimation assuming exchangeability $\widehat{U(t)}$ is represented by the dashed red line, and the estimation under the i.i.d. assumption $\widehat{U(t\mid \lambda)}$ by the blue dashed line.

\begin{figure}
	\centering
	\subfloat[]{\includegraphics[scale=.45]{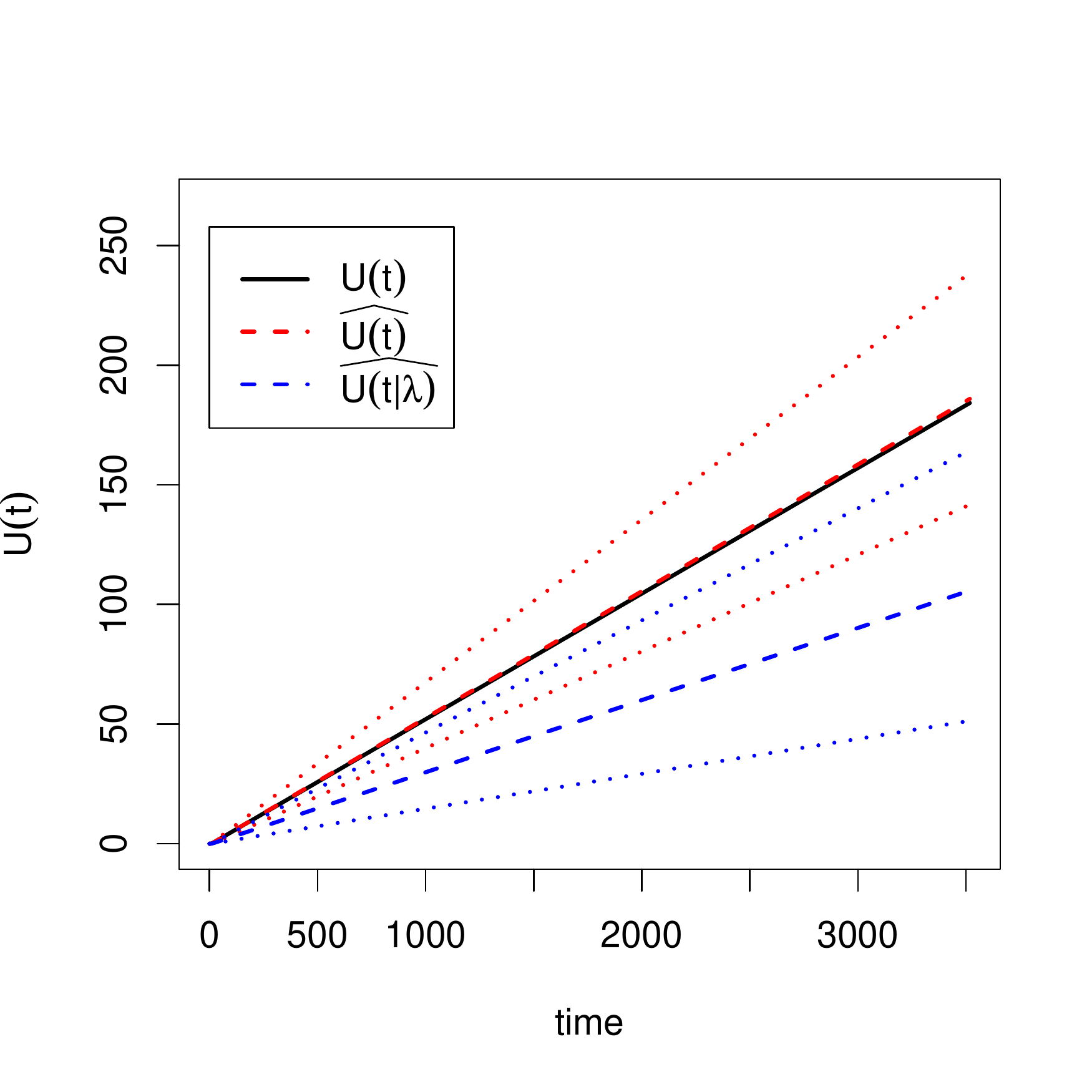}} 
	\subfloat[]{\includegraphics[scale=.45]{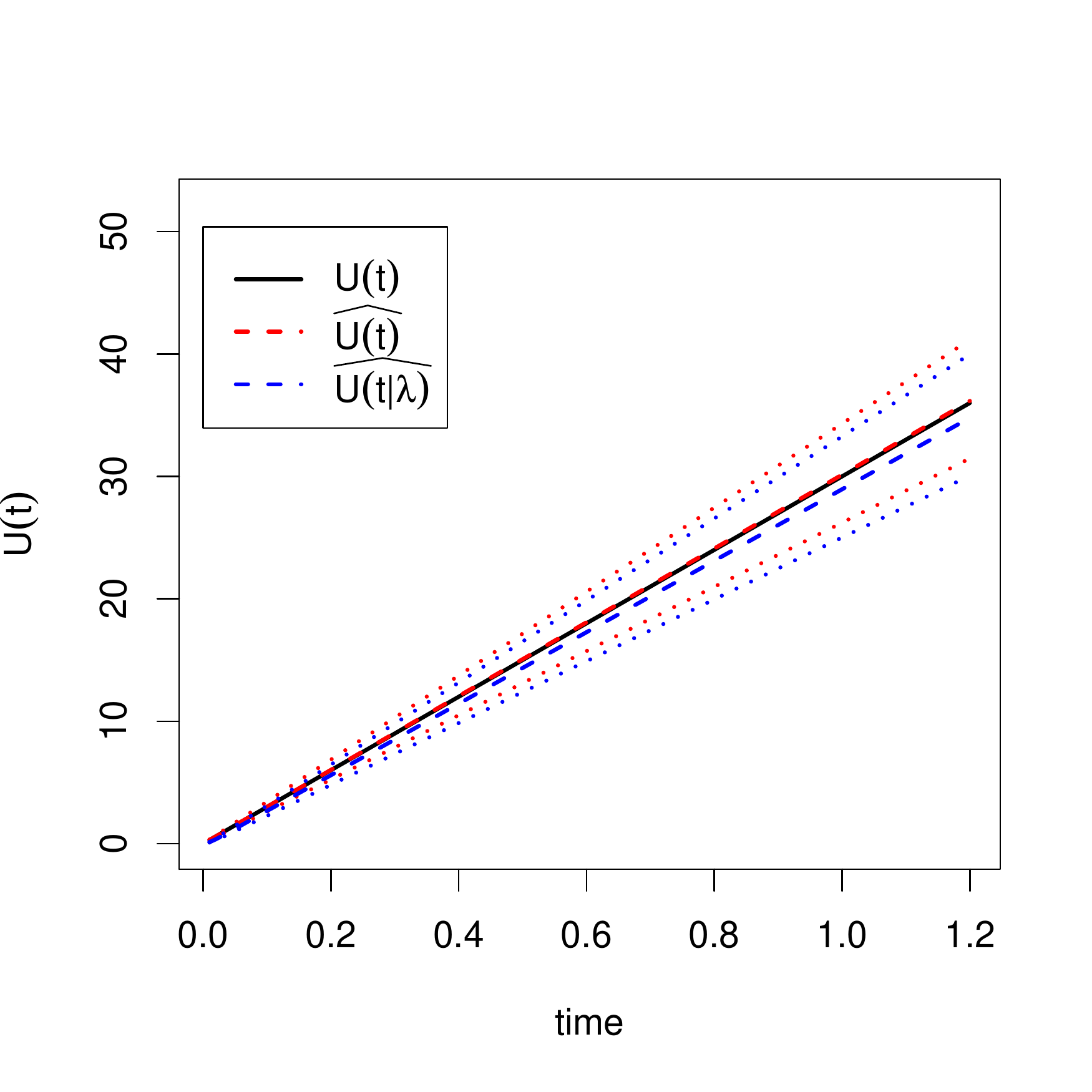}}  
	\caption{True renewal function (black line),  Monte Carlo estimations  (dashed lines) and percentile 2.5\% and 97.5\% of the estimations (dotted lines). Example 1 (a), Example 2 (b).}\label{fig:montecarlo}
\end{figure}

{Table} \ref{tab:examples} shows the true and the estimated values under the exchangeability assumption for the parameters in both examples. In general, the estimations are very close to the true values in these realizations. With the purpose of evaluating the uncertainty of the estimations, we performed a Monte Carlo simulation study. We consider 1,000 realizations of  Examples 1 and 2, and then performed the estimation of $U(t)$ using $\widehat{U(t)}$ and  $\widehat{U(t\mid \lambda)}$, the results are presented in the Figure \ref{fig:montecarlo}.  

\begin{table}[h] 
	\centering
	\begin{tabular}{@{} c|cc|cc @{}}
		&  \multicolumn{2}{c|}{ Example 1} & \multicolumn{2}{c}{ Example 2}  \\ 
		\hline
		Parameter & True & Estimated & True & Estimated \\ 
		\hline
		$\alpha$ & 2.1 & 2.3 & 30 & 31.7\\ 
		$m$ & 40 & 41 & 1 & 1\\ 
		${\mathsf{Corr}}(T_i, T_j)$ & 0.973 & 0.967 & 0.033 & 0.032\\ 
	\end{tabular}
	\caption{True and maximum likelihood estimation for the parameters in examples 1 and 2.}
	\label{tab:examples}
\end{table}

In both examples, the true renewal function is well estimated by the exchangeable model. As expected, the i.i.d. model only estimates well in Example 2 where the correlation is close to zero. From the Figures  \ref{fig:example12} and \ref{fig:montecarlo}, we conclude that when the dependence is ignored the renewal function is underestimated. 

\subsection{Reliability of hydraulic subsystems in load-haul-dump machines}

\begin{figure}[!h]
	\centering
	\includegraphics[scale=0.45]{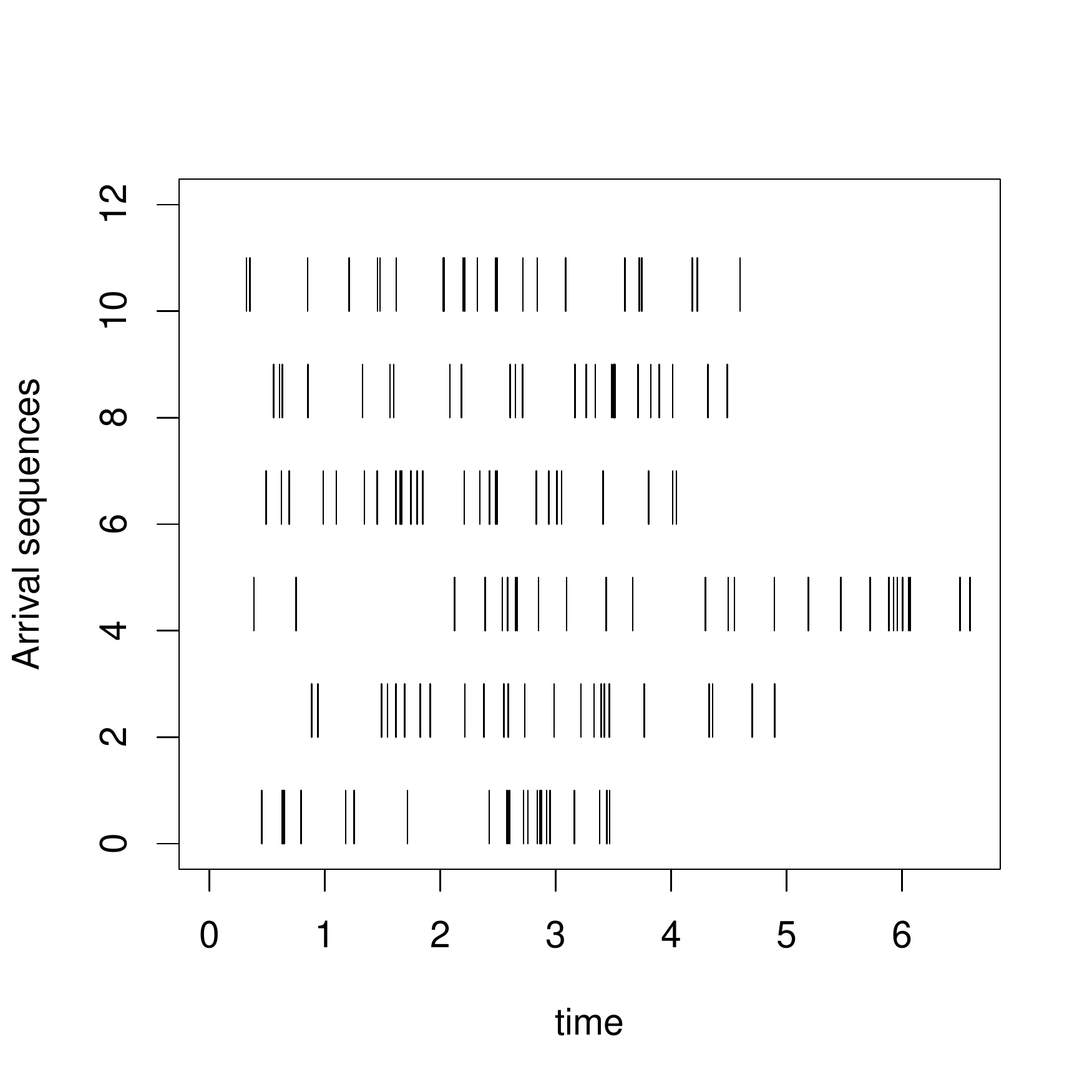}
	\caption{Arrival sequences of the times between successive failures of the hydraulic subsystems in LHD machines.}\label{fig:hydraulicdata}
\end{figure}

Here, we analyze the times between successive failures of the hydraulic subsystems in load-haul-dump (LHD) machines. The LHD machines are used in the mining industry to pick up ore or waste rock in the mines. The data consist of {inter-failure}  times of the hydraulic system for six LHD machines. The machines were identified as: LHD1 and LHD3 (old machines), LHD9 and LHD11 (medium old machines) and LHD17, LHD20, (new machines). Figure \ref{fig:hydraulicdata} displays the data.

\begin{figure}
	\centering
	\includegraphics[scale=0.45]{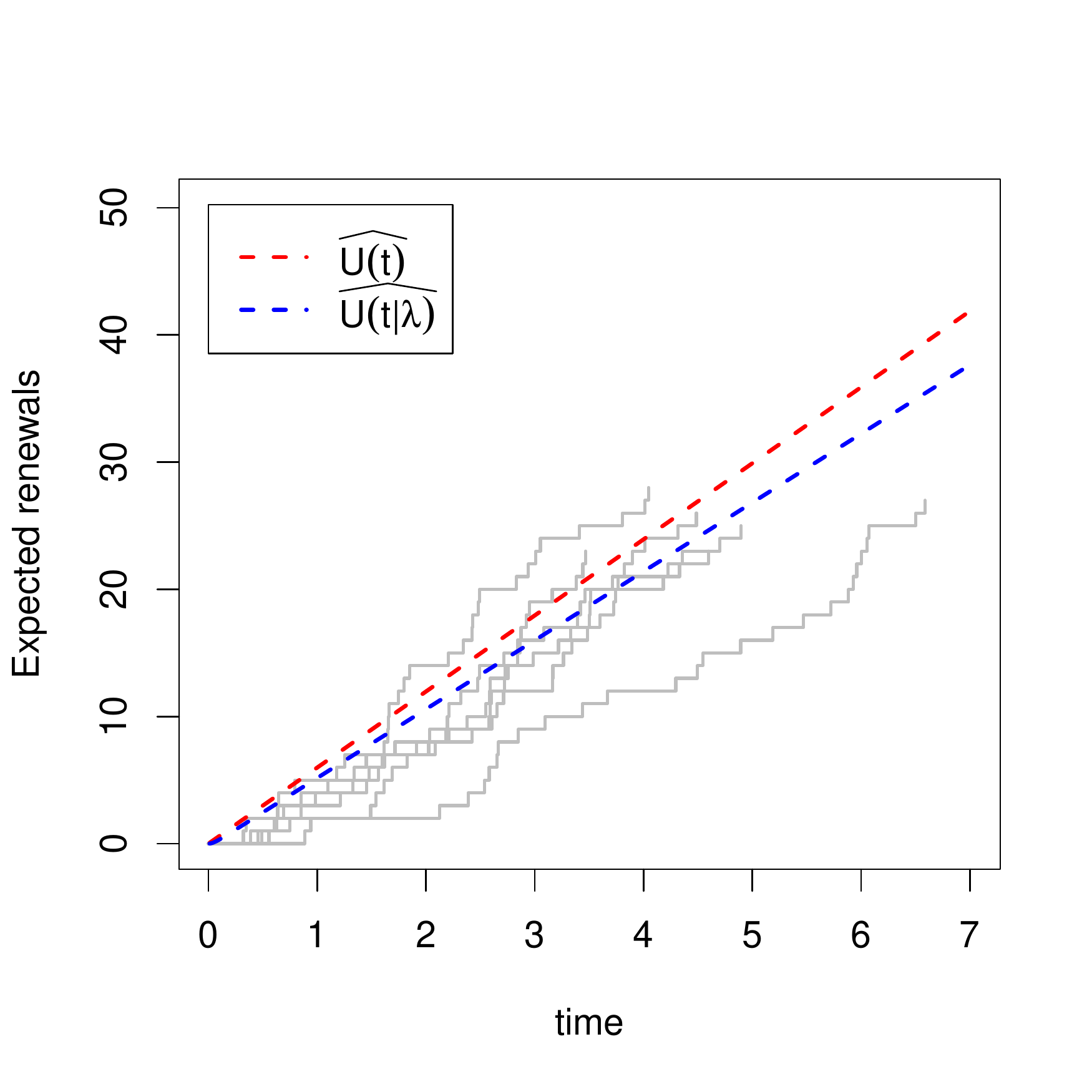}
	\caption{Trajectories (gray lines) and renewal function estimated under exchangeability (dashed red line) and under the i.i.d. assumption (dashed blue line). The time resolution is in months.}\label{fig:renewalHydraulic}
\end{figure}

These data, recollected and analyzed by \cite{Kumar92}, suggest that the i.i.d. assumption is not valid as, upon reparation, sometimes the machinery is not restored to a same-as-new condition. In this case, the values of the maximum likelihood estimation of the parameters were $\hat{m}=1$ and $\hat{\alpha}=5,982$. With the above values, the correlation is 0.167, which  {supports} the violation of the i.i.d. assumption. Figure \ref{fig:renewalHydraulic} compares the estimation of the renewal function under exchangeability and under the i.i.d. assumption. From this,  we can conclude that after three months of operation, we expect 18 failures for each machine under the exchangeable model and 16 failures under the i.i.d. model. As we concluded in the previous illustration, when dependence is ignored the renewal function is underestimated. The expected number of failures is a key element in the production lines, as it allows for a better planning, anticipating, for example, the number of replacement parts needed in the line. 

\section{Discussion}\label{sect:discussion}
{$\mu$-mixed renewal} processes represent an appealing alternative for renewal processes applications, where the sequence of failure times exhibit  dependence. The compromise when contrasted to the i.i.d. case is minimal. We have shown, via simulated and real scenarios,  how the resulting renewal functions can differ, leading thus to important differences such as those encountered in reliability engineering application. An important aspect to emphasize  is that the i.i.d. case is a particular case of exchangeability, thus nothing is lost when using the more general $\mu$-mixed renewal process. As a byproduct, Proposition~\ref{prop:nonparametric1} gives the distribution  sums of exchangeable random variables directed by the Dirichlet process. This  result is of interest in its own, e.g. for other studies and  limiting results for sums of exchangeable random variables (e.g. \cite{TaylorETAL86}).  Other choices of de Finetti's measures, potentially lead to similar outcomes, e.g. using the class of nonparametric priors resulting from normalizing completely random measures (see \cite{Lijoi2010}).  These, more general classes of de Finetti's measures, will be pursued elsewhere.

\section*{Acknowledgements}
The work of the first author was supported by CONACyT grant 241195 and DGAPA Posdoctoral Scholarship. The work of the second author was partially supported by ``Proyecto REDES ETAPA INICIAL, Convocatoria 2017 REDI170094" and by Millennium Science Initiative of the Ministry of Economy, Development, and Tourism, ``Millennium Nucleus Center for the Discovery of Structures in Complex Data". The work was concluded during a visit by the third author to the Department of Statistics \& Data Sciences at the University of Texas at Austin. Hospitality from the department is gratefully acknowledged as is support from a Fulbright Scholarship.

\bibliographystyle{tfnlm} 

\bibliography{Renewal_ArXiv}

\section*{Appendix}

Proof of Theorem \ref{theo:relationU(t)andU(t|F)}
\begin{proof}
	Conditioning over the values of the random distribution $ \Fvar $, we obtain: 
	\begin{equation*}
	U(t)=\esp{ N(t)} =\espmu{\esp{\left. N(t)\right|\Qrand}  }=\int_\clg{F} U(t|\Fvar)\mu(d\Fvar).
	\end{equation*}
	Where the last equality follows from {the de Finetti's representation theorem.}
\end{proof} 

Proof of Proposition \eqref{prop_U_sum_conv_F}

\begin{proof}
	Conditioning with $ \Fvar $ we obtain,
	\begin{align*}
	U(t)=&\espmu{\esp{N(t)|\Qrand}}\\[2ex]
	=&\int_\clg{F} \sum_{n=1}^{\infty}\Fvar^{*n}(t) \mu(d\Fvar)\\[2ex]
	=&\sum_{n=1}^{\infty}\int_\clg{F} \Fvar^{*n}(t) \mu(d\Fvar),
	\end{align*}
	in the third equality we use the fact that $ \esp{N(t)|\Qrand=\Fvar} $ is the renewal function in the i.i.d. case with renewals distributed $ \Fvar $, which can be expressed as $ \sum_{n=1}^{\infty}\Fvar^{*n}(t) $, see, for example, \cite{Gut_Stopped_random_walks_b09}.
\end{proof}

Proof of Proposition \eqref{prop_Laplace_ex_renewal}
\begin{proof}
	\begin{align*}
	L_U(s)=&\int_0^\infty e^{-st}U(t)dt
	=\int_0^\infty e^{-st}\sum_{n=1}^{\infty}\int_\clg{F} \Fvar^{*n}(t) \mu(d\Fvar)dt\\[2ex]
	=&\int_\clg{F} \sum_{n=1}^{\infty}[L_{\Fvar}(s)]^n \mu(d\Fvar)=\int_\clg{F} \dfrac{L_{\Fvar}(s)}{1-L_{\Fvar}(s)}\mu(d\Fvar),
	\end{align*}
	in the last equality we used the fact that  $ 0<L_{\Fvar}(s)<1. $
\end{proof}

Proof of Theorem \ref{thm_first_eq_of_U}

\begin{proof}
	Conditioning in $ T_1 $ and $ \Fvar $, we obtain
	\begin{align*}
	U(t)=&  \int_\clg{F} \int_0^\infty\esp{N(t)|T_1=x,\Qrand=\Fvar}d\Fvar(x)\mu(d\Fvar)\\[2ex]
	=&  \int_\clg{F} \int_0^t[1+U(t-x|\Fvar)]d\Fvar(x)\mu(d\Fvar)\\[2ex]
	=&  \int_\clg{F} \int_0^t d\Fvar(x)\mu(d\Fvar)+ \int_\clg{F} \int_0^tU(t-x|\Fvar)d\Fvar(x)\mu(d\Fvar)\\[2ex]
	=&  \espmu{\Qrand(t)}+ \espmu{\Qrand*U(t|\Qrand)}.
	\end{align*}	
\end{proof}

\bigskip

Proof of Theorem \ref{thm_key_ex_rev_theorem}
\begin{proof}\label{proof_thm_key_ex_rev_theorem}
	The function
	\begin{equation*}
	A'(t|\clg{G})=a(t)+ \espmu{a*U(t|\Fvar);\Fvar\in\clg{G}},  \qquad t\ge0,\clg{G}\in\mathfrak{F},
	\end{equation*}
	is bounded over bounded intervals for fixed $  \clg{G}\in\mathfrak{F}$. For a fixed $ T\in\mb{R}_+ $, we have
	\begin{align*}
	\sup_{0\le t\le T}A'(t|\clg{G})=&\sup_{0\le t\le T}\left(a(t)+ \int_\clg{G} a*U(t|\Fvar)\mu(d\Fvar) \right)\\[2ex]
	\le &\sup_{0\le t\le T}a(t)+ \int_\clg{G} \sup_{0\le t\le T}	\int_0^ta(t-x)dU(x|\Fvar)\mu(d\Fvar) \\[2ex]
	\le &\sup_{0\le t\le T}a(t)+ \int_\clg{G} \int_0^T\sup_{0\le t\le T}	a(t)dU(x|\Fvar)\mu(d\Fvar) \\[2ex]
	=&\sup_{0\le t\le T}a(t)+ \sup_{0\le t\le T}	a(t)\int_\clg{G} \int_0^TdU(x|\Fvar)\mu(d\Fvar) \\[2ex]
	\le&\sup_{0\le t\le T}a(t)\left(1+\int U(T|\Fvar)\mu(d\Fvar) \right)\\[2ex]
	=&\sup_{0\le t\le T}a(t)\left(1+U(T) \right).
	\end{align*}
	
	This implies that \[ \sup_{0\le t\le T \enspace\clg{G}\in\clg{Q}
	}A'(t|\clg{G})<\infty. \]
	
	Let see that $ A' $ accomplishes \eqref{eq_renewal}. For the definition of $ A' $, we see that 
	\begin{align*}
	A'(t|\clg{F})=& a(t)+\espmu{a*U(t|\Fvar)},  \\[2ex]
	A'(t|\Fvar)=&  a(t)+a*U(t|\Fvar) =  a(t)+a*\sum_{n=1}^{\infty}\Fvar^{*n}(t). 
	\end{align*}
	
	These last expressions implies:
	\begin{align*}
	A'(t|\clg{G})=& a(t)+\espmu{a*U(t|\Fvar);\Fvar\in\clg{G}}  \\[2ex]
	=& a(t)+\int_\clg{G} a*\left(\Fvar(t)+\Fvar*\sum_{n=1}^{\infty}\Fvar^{*n}(t)\right) \mu(d\Fvar) \\[2ex]
	=& a(t)+\int_\clg{G} \Fvar*\left(a(t)+a*\sum_{n=1}^{\infty}\Fvar^{*n}(t)\right) \mu(d\Fvar) \\[2ex]
	=&  a(t)+ \espmu{\Fvar*A'(t|\Fvar);\Fvar\in\clg{G}}.
	\end{align*}
	
	Therefore, $ A' $ fulfills \eqref{eq_renewal}. To finish the proof, we have to prove that the solution is unique.  Assume that the function $ B$ is bounded over intervals, and it is also a solution to (\ref{eq_renewal}), i.e.,
	\[ B(t|\clg{G}) =a(t)+\espmu{\Fvar*B(t|\Fvar);\Fvar\in\clg{G}}.\]

	As already mentioned the evaluation on $ \Fvar $ and $ \clg{F}$ plays an important role. In the case of \eqref{eq_renewal} we obtain that
	\begin{align*}
	A(t|\clg{F})=& a(t)+\espmu{\Fvar*A(t|\Fvar)},  \\[2ex]
	A(t|\Fvar)=&  a(t)+\Fvar*A(t|\Fvar),  
	\end{align*}
	whit analogously for $ B(t|\clg{F})$ and $ B(t|\Fvar) $. These last expressions imply
	\begin{align*}
	|A(t|\Fvar)-B(t|\Fvar)|=& |\Fvar*[A(t|\Fvar)-B(t|\Fvar)]|\\[2ex]
	=& |\Fvar^{*n}*[A(t|\Fvar)-B(t|\Fvar)]|\\[2ex]
	\le&  \Fvar^{*n}(t)  \sup_{0\le s\le t, \clg{G}\in\mathfrak{F} }|A(s|\clg{G})-B(s|\clg{G})|, 
	\end{align*} and so
	\begin{align*}
	|A(t|\clg{G})-B(t|\clg{G})|=& \left|\espmu{\Fvar*A(t|\Fvar)-\Fvar*B(t|\Fvar);\Fvar\in\clg{G}}\right| \\[2ex]
	\le&  \espmu{|\Fvar*[A(t|\Fvar)-B(t|\Fvar)]|;\Fvar\in\clg{G}} \\[2ex]
	\le& \espmu{\Fvar^{*n}(t) ;\Fvar\in\clg{G}}  \sup_{0\le s\le t, \clg{G}\in\mathfrak{F}}|A(s|\clg{G})-B(s|\clg{G})|, 
	\end{align*}
	and we obtain the uniqueness using that $ A $ and $ B $ are bounded and $ \espmu{\Fvar^{*n}(t)}\to0 $ as $ n\to\infty$. This last argument completed the proof.
\end{proof}

Proof of Proposition \ref{prop:nonparametric1}

\begin{proof} 
	Let $V=(v_1,\ldots,v_n)$  be the random vector indicating the repeated values in $(T_1,\ldots,T_n)$, i.e. there are $v_1$ values that only repeats once, $v_2$ values that repeats twice, and so on. 	
	Then, conditioning on $V$ the distribution of the adding process can be written as 
	\[ \prob{T_1+\ldots+T_n\le t} = \sum_v \prob{T_1+\ldots+T_n\le t|V=v}\prob{V=v}. \]
	
	The conditional distribution of $ T_1+\ldots+T_n $ given $ V=v $ is equal to the convolution of $ v_1 $ independent variables with distribution $ H_{1} $, convoluted with $ v_2 $ variables distributed $ H_2 $, and so on. The distributions $ H_j $ are needed because the same value is repeated $ j $ times, and we need the probability  $ \prob{jT\le t} $, for $ T\sim\dist{H} $. Thus,
	\[ \prob{T_1+\ldots+T_n\le t|V=v} = \left( H_1^{*v_1}\ast H_2^{*v_1}\ast\cdots\ast H_n^{*v_n}\right) (t), \]
	Finally, $ \prob{V=v} $ is given by the Ewen's sampling formula  \cite{Ewen_a72} 
	\[ \prob{V=v} = \dfrac{n!}{(\alpha)_n}\prod_{j=1}^n \dfrac{\alpha^{v_j}}{j^{v_j}v_j!}, \]
	
	\noindent which completes the proof of (i). For (ii), we apply {{Proposition \ref{prop_U_sum_conv_F}}.} 
\end{proof} 

\end{document}